\documentclass[showpacs,prb,onecolumn,preprint]{revtex4}
\usepackage{amssymb}
\usepackage{amsmath}

\input{tcilatex}

\begin{document}

\title{Controllable quantum spin precession by Aharonov-Casher phase in
conducting ring }
\author{Shun-Qing Shen}
\affiliation{Department of Physics, The University of Hong Kong, Pukfulam Road, Hong
Kong, China}
\author{Zhi-Jian Li}
\affiliation{Institute of Physics, Chinese Academy of Sciences, Beijing 100080, China}
\author{Zhongshui Ma}
\affiliation{State Key Laboratory for Mesoscopic Physics and Department of Physics,
Peking University, Beijing 100871, China}
\date{June 26, 2003}

\begin{abstract}
We investigate quantum spin transport in a structure of conducting ring,
embedded in textured electric field, with two leads, and obtain an exact
solution for the problem. The spin precession induced by the Aharonov-Casher
phase is studied. It is shown that the spin-polarized current and its
polarizability can be controlled by the electric field. As a result the
polarizability is a function of the geometric phase which originates from
the spin-orbital interaction in the ring.
\end{abstract}

\maketitle

In the recent years spin transport has attracted considerable interests
because of its potential applications in semiconductor electronics and
quantum computation \cite{1,2,3}. How to control or to modulate spin
polarized current is one of important steps in the investigation of spin
coherence in electronic systems. Datta and Das \cite{datta} proposed a spin
field-effect transistor (FET), in which spin polarization of charge carriers
precesses when the charge carriers transmit through a 2D semiconductor
channel between a ferromagnetic spin injector and a ferromagnetic spin
collector. It is well known that, in semiconductor heterostructures, the
spin-orbit or Rashba interaction makes spin-orbit splitting of the
conduction electron energy band. Such splitting produces spin precession
while the electrons go through the semiconductor channel. Experimentally,
the Rashba coupling can be adjusted by an electric field such that a
spin-polarized current injected from the source can be spin-dependently
modulated on its way to the drain by an external gate \cite{nitta,Grundler00}%
. Other proposals were recently reported \cite{4,5,6,7,8}. However more
sophisticated structures are conceivable where current modulation could
arise from the spin interference.

The conventional Aharonov-Bohm (AB)\ effect is anticipated to modulate the
charge current. It was shown that a non-uniform magnetic field may also
control the polarization direction of spin current \cite{Frustaglia01}. The
Aharonov-Casher (AC) effect \cite{ac} originates from the spin-orbit
coupling between the moving magnetic polar and the electric field. It is
expected to implement the spin current modulation manifested by AC effect.
Quantum coherent transport in a structure of the AC ring with two leads was
discussed by many authors \cite{Mathur92, Aronov, su, choi, Nitta99}. It was
found that the AC\ flux can lead to interference phenomena as the AB flux
with an observable AC oscillation in the conductance. With an electric field
as a parameter, it is expected to study the precession of spin current for
interfering electrons manifested by the geometric properties of the ring. In
the present paper we establish the connection between quantum spin
polarizability and spin precession in the quantum transport in a AC ring,
such that it is controllable through the external electric field by means of
AC effect. We present an exact solution for the problem and focus on the
spin aspect of quantum transport. Assume the ring embeds in a symmetric
textured electric field with an arbitrary tilt angle as shown in Fig. 1. In
the AC ring, the magnetic moment of charge carriers is influenced by the
electric field through the spin-orbit interaction and there is an energy
splitting between spin-up and -down electrons. The local spin orientations
on the ring are determined by a spin cyclic evolution over the ring. When
electrons travel through each arm of the ring, the wave functions of
electrons acquires spin-dependent Aharonov-Anandan (AA) phases \cite%
{Aharonov87}, which is the geometric part of the AC phase. Without
specializing the direction of spin-polarized injection, we calculate the
spin-dependent transmission coefficients analytically. In the Landauer
framework of ballistic transport \cite{land}, it is shown that the spin
polarized current and its polarizability can be controlled by the electric
field via AC phase. Our result shows that the AC ring can act as a tunable
spin switch by adjusting the electric field.

In the ring, the one-particle Hamiltonian for non-interacting electrons with
a momentum $\mathbf{p}$ is given by 
\begin{equation}
H=\frac{1}{2m}(\mathbf{p}-\frac{\mu }{2c}\mathbf{\sigma }\times \mathbf{E}%
)^{2},
\end{equation}%
where $\mu =g\mu _{B}$ a magnetic moment of charge carrier, $\mu _{B\text{ }%
} $the Bohr magneton, $g$ the gyromagnetic factor, $c$ the velocity of light
in vacuum and $\sigma _{i}$ ($i=1,2,3$) the Pauli matrices. The linear term
of the electric field in the Hamiltonian, $\mathbf{\sigma \cdot }\left( 
\mathbf{E}\times \mathbf{p}\right) ,$ represents the spin-orbit coupling and
is proportional to the electric field. In the cylindrical coordinate, the
textured electric field can be written as $\mathbf{E}=E(\cos \chi \mathbf{%
\hat{r}-\sin \chi \hat{z}})$. The eigen functions of given Hamiltonian (1)
are given by $\Psi _{n,\pm }=\frac{1}{\sqrt{2\pi }}e^{in\phi }\xi _{\pm }$
where 
\begin{eqnarray}
\xi _{+} &\equiv &\left( \cos \frac{\beta }{2},e^{i\phi }\sin \frac{\beta }{2%
}\right) ^{T}\text{,}  \label{wave1} \\
\xi _{-} &\equiv &\left( \sin \frac{\beta }{2},-e^{i\phi }\cos \frac{\beta }{%
2}\right) ^{T}.  \label{wave2}
\end{eqnarray}%
where T stands for the transpose and $\pm $ denotes spin up and down along
the direction of unit vector $\Omega =(\sin \beta \cos \phi ,\sin \beta \sin
\phi ,\cos \beta ).$ $\beta $ is defined by $\tan \beta \equiv \omega
_{1}/(1+\omega _{3})$ with $\omega _{1}\equiv (\mu Ea/\hbar c)\sin \chi $
and $\omega _{3}\equiv (\mu Ea/\hbar c)\cos \chi $. The corresponding
eigenvalues are $E_{n,\pm }=\frac{\hbar ^{2}}{2ma^{2}}\left( n-\Phi
_{AC}^{\pm }/2\pi \right) ^{2},$ where $n$ is an integer. $\Phi _{AC}^{\pm
}=-\pi \pm \phi _{0}$ ($\phi _{0}=\pi \sqrt{\omega _{1}^{2}+(1+\omega
_{3})^{2}}$) are the AC phases which are acquired while the two spin states
evolves in the presence of the electric field \cite{ac}.

Assume that electrons are free electrons in two leads and have the momentum $%
\hslash k.$ The energy is given by $E=\hbar ^{2}k^{2}/2m$. When an electron
is transported along one arm in the clockwise direction from the input
intersection A, it acquires an AC phase $\Phi _{AC}^{\pm }/2$ at the output
intersection B. And the electron acquires a phase $-\Phi _{AC}^{\pm }/2$ in
the counter-clockwise direction along the other arm. Thus the total phase
changes around the loop is $\Phi _{AC}^{\pm }$. In the ring the electric
field may change the momenta of the electrons into the same energy as in the
leads in two different spin states $\xi _{\pm }$, $k_{1}^{\pm }=k+\Phi
_{AC}^{\pm }/2\pi a$ and $k_{2}^{\pm }=k-\Phi _{AC}^{\pm }/2\pi a,$ where
the subscripts $1,2$ denote the clockwise and counter-clockwise direction,
respectively. $a$ is the radius of the ring. It is worth to point out that $%
\left( k_{1}^{\pm }-k_{2}^{\pm }\right) \pi a=\Phi _{AC}^{\pm }$ is
independent of the momentum of incident particles. This phase difference
leads two branches of wave functions to intervene at the output intersection
($x$ ranges from $0$ to $\pi a$ in two arms of the ring). The wave function
in two arms of the ring can be written as 
\begin{eqnarray}
\Psi _{1} &=&\sum_{\alpha =\pm }\left( c_{1,\alpha }e^{ikx}+d_{1,\alpha
}e^{-ikx}\right) e^{i\Phi _{AC}^{\alpha }x/2\pi a}\xi _{\alpha }\left( +%
\frac{x}{a}\right) ;  \label{arm-1} \\
\Psi _{2} &=&\sum_{\alpha =\pm }\left( c_{2,\alpha }e^{ikx}+d_{2,\alpha
}e^{-ikx}\right) e^{-i\Phi _{AC}^{\alpha }x/2\pi a}\xi _{\alpha }\left( -%
\frac{x}{a}\right) .  \label{arm-2}
\end{eqnarray}%
We have assumed that a polarized electron is injected to left electrode from
ferromagnetic source and travel in the $x$ direction, and the spin state of
the injected electron is $\Psi _{i}=\left( \cos \alpha ,\sin \alpha \right)
^{T}e^{ikx}.$ For a general considerations the wave functions in the two
arms can be expressed in terms of the reflection and transmission of
electrons when they transport through the ring in presence of the
spin-orbital interaction. The wave functions of the electron in the left and
right electrodes are 
\begin{eqnarray}
\Psi _{l} &=&\Psi _{i}+\left( 
\begin{array}{c}
r_{\uparrow } \\ 
r_{\downarrow }%
\end{array}%
\right) e^{-ikx}, \\
\Psi _{r} &=&\left( 
\begin{array}{c}
t_{\uparrow } \\ 
t_{\downarrow }%
\end{array}%
\right) e^{ikx},
\end{eqnarray}%
where $r_{\sigma }$ and $t_{\sigma }$ are the spin dependent reflection and
transmission coefficients, respectively.

In the linear response regime of the Landauer framework \cite{land}, the
quantum mechanical transmission amplitude is related to two-probe
conductance. The spin-dependent conductance through the ring is expressed in
term of the transmission probability $T_{\alpha }=\left\vert t_{\alpha
}\right\vert ^{2}$ as $G_{\alpha }=(e^{2}/h)T_{\alpha }.$ To calculate the
transmission probability through the AC ring, wee use the local coordinate
system for each circuit such that the $x$ coordinate is taken along the
electron current.\cite{xia,choi} The origin of each local coordinate is
taken at each intersection. The choice of the coordinate origin is trivial
because it only affects an extra phase factor on the transmission amplitude.
Thus the Griffith's boundary condition \cite{griffith} gives that at each
intersection the wave function is continuous and the current density is
conserved. After\ some tedious algebra, we obtain the transmission
coefficient $t$: 
\begin{eqnarray}
t &=&\frac{i\sin k\pi a\sin \frac{\phi }{2}}{\sin ^{2}\frac{\phi }{2}-\left(
\cos k\pi a-\frac{i}{2}\sin k\pi a\right) ^{2}}\sigma _{z}\gamma \sigma
_{z}\gamma \left. \Psi _{i}\right\vert _{x=0}  \notag \\
&=&\frac{i\sin k\pi a\sin \frac{\phi }{2}}{\sin ^{2}\frac{\phi }{2}-\left(
\cos k\pi a-\frac{i}{2}\sin k\pi a\right) ^{2}}\left( 
\begin{array}{c}
\cos \left( \alpha -\beta \right) \\ 
\sin \left( \alpha -\beta \right)%
\end{array}%
\right)  \label{tc}
\end{eqnarray}%
where $\gamma =\cos \frac{\beta }{2}\sigma _{z}+\sin \frac{\beta }{2}\sigma
_{x}.$ The corresponding transmission coefficient is given by 
\begin{equation}
T=\left\vert t_{\uparrow }\right\vert ^{2}+\left\vert t_{\downarrow
}\right\vert ^{2}=\left\vert \frac{\sin k\pi a\sin \frac{\phi }{2}}{\sin ^{2}%
\frac{\phi }{2}-\left( \cos k\pi a-\frac{i}{2}\sin k\pi a\right) ^{2}}%
\right\vert ^{2}.  \label{t}
\end{equation}%
The coefficient probability was obtained for two special cases by Choi et
al. \cite{choi} For instance when $\alpha =\beta /2,$ $c_{n,-}=d_{n,-}=0$.
In this sense only the component of $\xi _{+}$ remains and there is no spin
flip. Our general expression shows that the transmission probability is
independent of the incident spin state. Both expressions in Eqs. (\ref{tc})
and (\ref{t}) are determined by the AC phase, the spin and kinetic state of
the incident particles, and the electric field. It is noticed that there is
a special case that the transmission coefficients become zero and all
incident wave reflects from the ring when $\sin k\pi a=0$. To illustrate the
main physics in the problem, we present some numerical calculation for
several physical quantities. For an InAs ring with $a=1\mu m$, the Fermi
velocity $v_{F}$ is approximately $3\times 10^{7}$cms$^{-1}$ and $ka\approx
60$\cite{su}. It is interesting that the transmission coefficient is only
determined by the decimal part of $ka$. The number is tunable for k and the
size of the ring. Thus in the numerical calculation we take $ka=60.239$.

Following the Landauer formula, the spin-resolved current is given by $%
j_{\sigma }=VG_{\sigma }=V\left( e^{2}/h\right) T_{\sigma }.$ One of the
interesting observations is that 
\begin{equation}
j_{e}\equiv j_{\uparrow }+j_{\downarrow }=V\frac{e^{2}}{h}T,
\end{equation}%
which is determined by the AC phase as well as the radius of the ring. It
depends on the energy of incident particles, but is independent of the
initial spin state. Thus the electric field can modulate the transmission
charge current through the AC\ phase. In Fig. 2, we plot $j_{e}$ versus the
magnitude $E$ and direction angle $\chi $ of the electric field. For a
specific size $a$ and an incident momentum $k$,\ we find that the
transmission charge current oscillates with the electric field for an
arbitrary $\chi $. From two explicit expressions for $t_{\uparrow }$ and $%
t_{\downarrow }$ we observe that, after going through the AC ring the
particles evolves from the spin state $\left( \cos \alpha ,\sin \alpha
\right) $ into a new one $\left( \cos \left( \alpha -\beta \right) ,\sin
\left( \alpha -\beta \right) \right) $. Thus the spin current along the S$%
_{z}$ axis is given by 
\begin{equation}
j_{s}\equiv \frac{\mu _{B}}{e}\left( j_{\uparrow }-j_{\downarrow }\right) =V%
\frac{e\mu _{B}}{h}T\cos 2\left( \alpha -\beta \right) .
\end{equation}%
In Fig.3, we plot the polarized spin current versus the electric field
magnitude $E$ and direction angle $\chi .$ The transmission polarized spin
current is controlled\ by the tilt angle $\beta $ as well as the modulation
analogous to the transmission charge current.

To see this modulation clear, we introduce a dimensionless polarizability $P$
to describe the polarization degree along the spin S$_{z}$ axis of the
transmitted current through the AC ring, which is defined by 
\begin{equation}
P\equiv \frac{j_{\uparrow }-j_{\downarrow }}{j_{\uparrow }+j_{\downarrow }}%
=\cos 2\left( \alpha -\beta \right) .  \label{p}
\end{equation}%
$P$ is independent of the mass $m$ and momentum $\hbar k$ of the incident
charge carriers for a large $ka$. This is an important advantage for the
device application. $P$ is similar to the \textit{spin injection rate}
defined in ferromagnetic/semiconductor/ferromagnetic heterostructures \cite%
{Johnson98}, and can be measured experimentally. Thus $P$ is simply
determined by $\alpha -\beta $ only. For\ the incident state $\Psi _{i}$,
the spin polarized tilt angle is $2\alpha $ and the polarizability $%
P=P_{0}\equiv \cos 2\alpha $. We see that the transmitted polarizability is
modulated by $\beta $. In the case the transmitted current is just modulated
by the AC\ phase, and is spin-independent. We plot the spin polarizability $%
P $ versus $E$ and $\chi $ in Fig.4 for the case $\alpha =0$. From the
definition of $\beta $ it is equal to $\chi $ when the electric field $E$ is
very large. It is obvious that the spin polarizability approaches to a
constant with increasing of electric field strength for a certain angle.
This mechanism is different from Datta and Das's proposal, where is the
phase difference of two spin states determines spin precession. In the
present paper, the role of phase difference $\left( k_{1}^{\pm }-k_{1}^{\pm
}\right) \pi a=\Phi _{AC}^{\pm }$ of two spin eigenstates between the source
and drain just controls the transmission coefficients and does not affect
the spin polarizability.

This theory can be applied to a simple device to control the polarized spin
current following Datta and Das' proposal. The semiconductor hetrostructure
is replaced by a conducting ring to connect the source and drain. The gate
voltage is applied to control the electric field exerting on the ring as
shown in Fig. 1(b). In this case $\mathbf{E}=-E\hat{z}$, i.e., $\chi =\pi /2$%
. Assume a spin-up electron ($\alpha =0)$ is injected from ferromagnetic
source and passes though the AC ring. The tilt angle is simply written as $%
\beta =\arctan (\mu Ea/\hbar c).$ The polarization direction of the
transmission current can be tuned by the electric field explicitly, i.e., 
\begin{equation}
P=\frac{1-\left( \mu Ea/\hbar c\right) ^{2}}{1+\left( \mu Ea/\hbar c\right)
^{2}}.
\end{equation}
For a large electric field $E\gg \hbar c/\mu a,$ $\beta \rightarrow \pi /2$
and $P=-1.$ The polarization direction of incident electron with spin up can
be flipped into spin down. \ One of the advantages of the device is either
the transmission probability or polarizability of the transmitted current is
tunable by the electric field. A similar device was proposed by Nitta et al. 
\cite{Nitta99}. Except for the electric field, an magnetic field is also
applied to split energy levels in the ring. According to the present theory
the electron spin can evolve a tilt angel even in absence of an external
magnetic field.

In summary, we have obtained an exact solution for the quantum transport in
a AC ring with two leads and shown that the quantum spin transport is
modulated by the electric field. The AC\ phase may control the transmission
current, and the spin of electron evolves an tilt angle $\beta $ after
passing through the ring. The angle is determined by the electric field and
the size of the ring, but independent of the energy of incident particles.
Following Datta and Das, we propose that the device of AC ring with two
leads can serve as a spin transistor. The spin-resolved current can be well
controlled by the voltage gate.

This work was supported by a grant from Research Grant Council of Hong Kong,
a CRCG grant of The University of Hong Kong, and the NNSFC-China. One of the
authors (SQS) acknowledge Max Planck Institute for Physics of Complex
Systems at Dresden, Germany for its support and hospitality, where part of
the work was done.

\FRAME{ftbpFU}{3.1427in}{0.9712in}{0pt}{\Qcb{(a) A symmetric textured
electric field excerted on the AC\ ring. The arrows indicates the incident,
transmitted and reflected current flows. (b) Spin transistor geometry: a AC
ring with two leads connecting to ferromagnetic source and drain is
sandwiched between two electrodes.}}{}{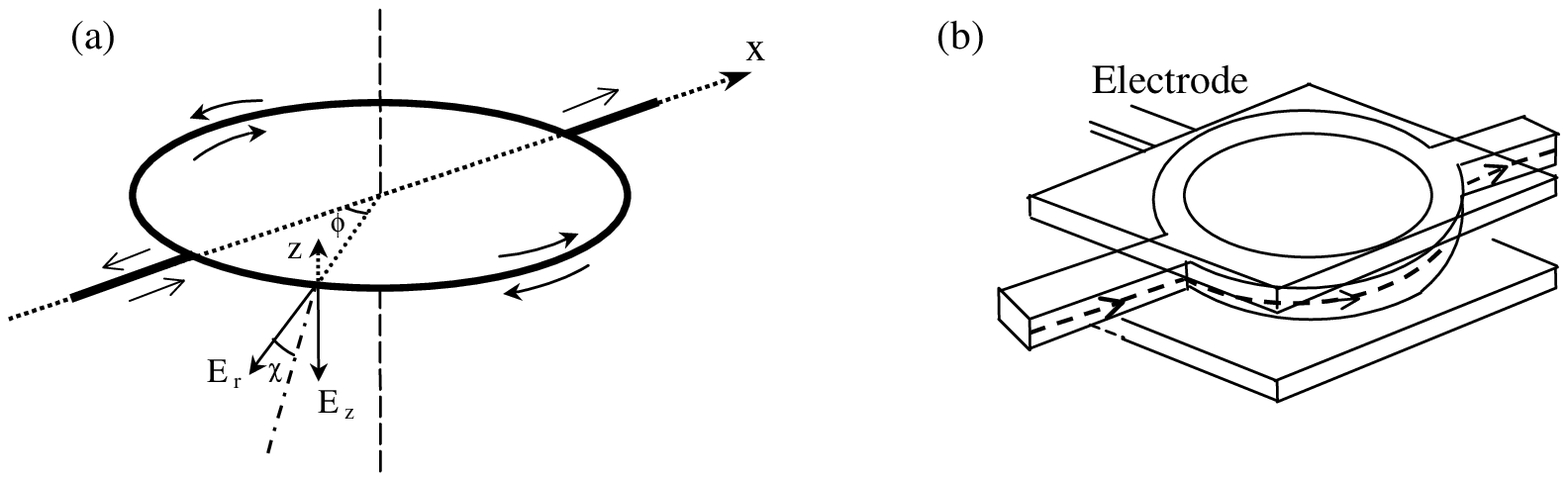}{\special{language "Scientific
Word";type "GRAPHIC";maintain-aspect-ratio TRUE;display "USEDEF";valid_file
"F";width 3.1427in;height 0.9712in;depth 0pt;original-width
6.186in;original-height 1.8922in;cropleft "0";croptop "1";cropright
"1";cropbottom "0";filename 'x1.eps';file-properties "XNPEU";}}

\FRAME{ftbpFU}{2.9983in}{2.2502in}{0pt}{\Qcb{The charge current $j_{e}$ with
unit $Ve^{2}/h$ varies with respect to the electric field $E$ with unit $%
\hbar c/\protect\mu a$ and its direction tilt angle $\protect\chi $ with
unit $\protect\pi $.}}{}{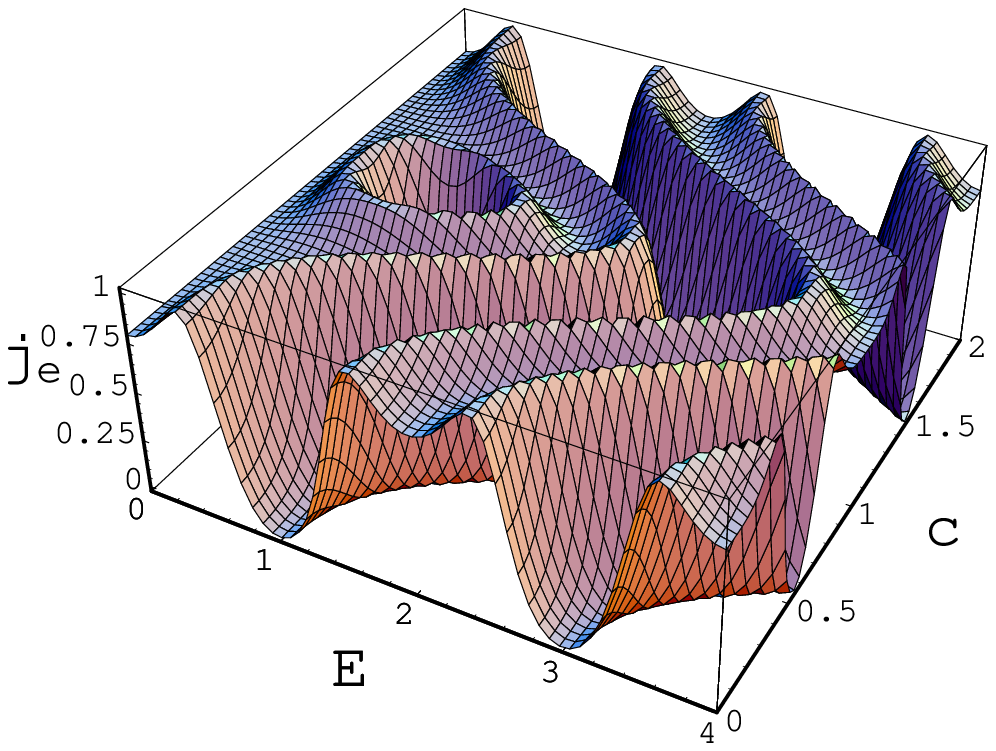}{\special{language "Scientific Word";type
"GRAPHIC";maintain-aspect-ratio TRUE;display "USEDEF";valid_file "F";width
2.9983in;height 2.2502in;depth 0pt;original-width 3.9972in;original-height
2.9922in;cropleft "0";croptop "1";cropright "1";cropbottom "0";filename
'x2.eps';file-properties "XNPEU";}}

\FRAME{ftbpFU}{3.0191in}{2.2658in}{0pt}{\Qcb{The polarized spin current $%
j_{s}$ with unit $Ve\protect\mu _{B}/h$ varies with respect to the electric
field $E$ with unit $\hbar c/\protect\mu a$ and its direction tilt angle $%
\protect\chi $ with unit $\protect\pi $.}}{}{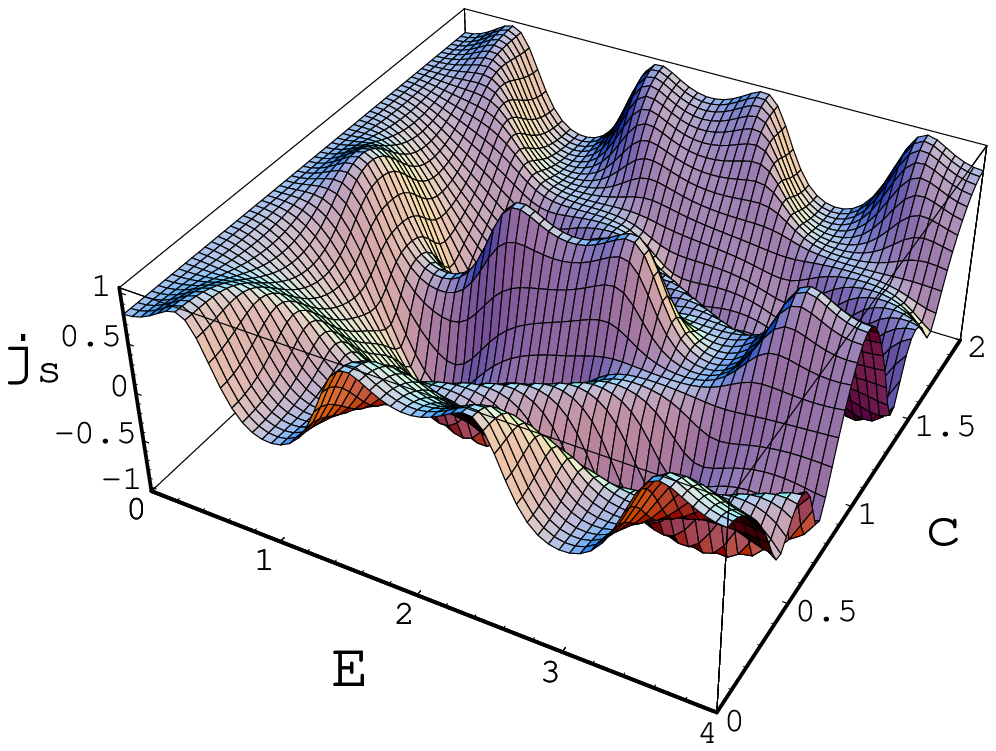}{\special{language
"Scientific Word";type "GRAPHIC";maintain-aspect-ratio TRUE;display
"USEDEF";valid_file "F";width 3.0191in;height 2.2658in;depth
0pt;original-width 3.9972in;original-height 2.9922in;cropleft "0";croptop
"1";cropright "1";cropbottom "0";filename 'x3.eps';file-properties "XNPEU";}}

\FRAME{ftbpFU}{3.0398in}{2.3436in}{0pt}{\Qcb{The spin polarizability of
current P with respect to the electric field $E$ with unit $\hbar c/\protect%
\mu a$ and its direction tilt angle $\protect\chi $ with unit $\protect\pi $.%
}}{}{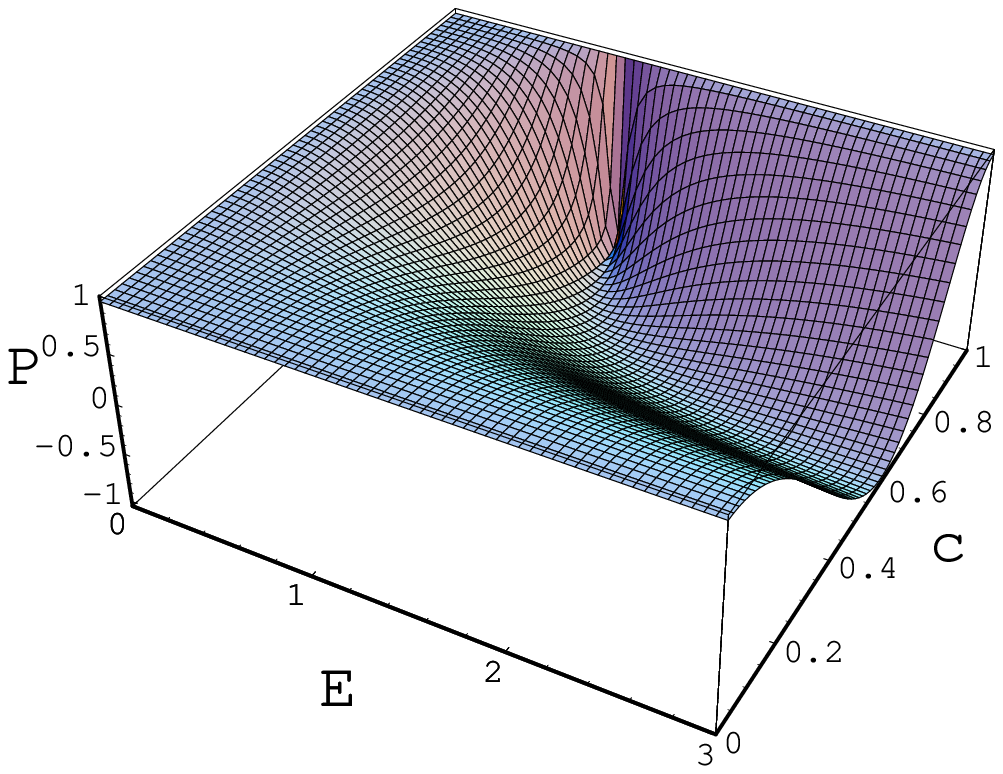}{\special{language "Scientific Word";type
"GRAPHIC";maintain-aspect-ratio TRUE;display "USEDEF";valid_file "F";width
3.0398in;height 2.3436in;depth 0pt;original-width 3.9972in;original-height
3.0753in;cropleft "0";croptop "1";cropright "1";cropbottom "0";filename
'x4.eps';file-properties "XNPEU";}}

\end{document}